\documentclass[twocolumn,prc,showpacs,preprintnumbers,superscriptaddress,floatfix]{revtex4}
\usepackage{dcolumn}
\usepackage{bm}
\usepackage{longtable}
\usepackage{graphicx,epsfig,latexsym,amssymb}
\usepackage{multirow,amsmath,array,booktabs,color}
\usepackage[section]{placeins}
\usepackage{mathrsfs}

\begin{document}

\title{General formalism of collective motion for any deformed system}
\author{Jian-You Guo }
\email[E-mail:]{jianyou@ahu.edu.cn}
\affiliation{School of physics and material science, Anhui University, Hefei 230601, P.R.
China}
\date{\today }

\begin{abstract}
Based on Bohr model, we have presented a general formalism describing the
collective motion for any deformed system, in which the collective
Hamiltonian is expressed as vibrations in the body-fixed frame, rotation of
whole system around the laboratory frame, and coupling between vibrations
and rotation. Under the condition of decoupling approximation, we have
derived the quantized Hamiltonian operator. Based on the operator, we have
calculated the rotational spectra for some special octupole and hexadecapole
deformed systems, and shown their dependencies on deformation. The result
indicates that the contribution of octupole or hexadecapole deformations to
the lowest band is regular, while that to higher bands is dramatic. These
features reflecting octupole and hexadecapole deformations are helpful to
recognize the properties of real nuclei with octupole and/or hexadecapole
deformations coexisting with quadrupole deformations.
\end{abstract}

\maketitle

\section{Introduction}

The theory of collective motion has been developed a long time ago. The
classical case corresponds to the quadruple deformations, which was
established by Bohr in 1952~\cite{Bohr521,Bohr522}. Bohr Hamiltonian is very
useful in describing the vibrations and rotation for quadruple deformed
nuclei. Especially for the shape evolution and phase transitions~\cite%
{Casten07}, Bohr Hamiltonian is a powerful tool to investigate the
critical-point symmetries like E(5)~\cite{Iachello00}, X(5)~\cite{Iachello01}%
, Y(5)~\cite{Iachello03}, and Z(5)~\cite{Bonatsos04}. More researches on the
collective motion by Bohr model can be found in the literatures~\cite%
{Casten06,Prochniak09,Cejnar10} and references therein. Several recent
progresses include Bohr Hamiltonian solved with a mass and deformation
dependent Kratzer potential~\cite{Bonatsos13}, an approximate analytical
formula for the energy spectrum for a prolate $\gamma $-rigid collective
Hamiltonian with a harmonic oscillator potential corrected by a sextic term~%
\cite{Budaca14}, and analytical solution of the Davydov-Chaban Hamiltonian
with a sextic potential for $\gamma =30^{\circ }$ and its satisfactory
description for the shape phase transition in Xe isotopes in comparison with
experiment~\cite{Buganu15}.

Bohr Hamiltonian is applicable to nuclei with quadruple deformations.
Although the quadruple deformations are the most frequently encountered in
real nuclei, the higher multipolar deformations are also essential for
satisfactory description of nuclear properties. The description of octupole
deformations has been a long-standing problem in nuclear physics~\cite%
{Butler96}. Theoretical calculations~\cite{Nazarewicz84,Nazarewicz92}
predicted the existence of octupole stable deformations and this problem
stirred considerable interest, especially in the Ce-Ba and the Rn-Th
regions. The level scheme of a few moderately or weakly deformed nuclei,
such as $^{64}\mathrm{Ge}$~\cite{Ennis91},$^{148}\mathrm{Sm}$~\cite{Urban91}%
, or $^{233,235}\mathrm{Ra}$~\cite{Chasman86} presents features that may be
related to octupole instabilities and softness of the nucleus with respect
to possible exotic octupole deformations. There has been evidence for the
existence of stable octupole deformations in the Rn-Th region~\cite%
{Cocks97,Cocks99}. For example, the existence of stable octupole
deformations in $^{224}$Ra has been verified in recent experiment~\cite%
{Gaffney13}. Furthermore, in the region $N=92,94$, octupole correlations
were observed in $^{150,152}$Ce isotopes~\cite{Zhu12,Li12}.

These show that there exist certainly the octupole deformations and/or
correlations in certain regions. For the study of collective motion
involving octupole deformations, the generalization of Bohr Hamiltonian was
explored in Ref.~\cite{Rohozinski82}. Its application to the problem of
octupole vibrations in nuclei was elaborated in the review~\cite%
{Rohozinski88}. The vibrational and rotational spectra obtained by the model
reproduce well the experimental data for some rare-earth and actinide nuclei~%
\cite{Dzyublik93,Denisov95}. In Refs.~\cite{Bonatsos05,Lenis06}, the
analytic solutions of Bohr Hamiltonian involving axially symmetric
quadrupole and octupole deformations with an infinite well potential or
Davidson potential were obtained, and normalized spectra and B(EL) ratios
were found to agree with experimental data for $^{226}$Ra and $^{226}$Th. As
there is the difficult to determine the intrinsic frame, the
parameterizations of octupole deformations were probed in Refs.~\cite%
{Rohozinski88,Rohozinski90,Hamamoto91,Wexler99}. Moreover, an alternative
parameterizations describing nuclear quadruple and octupole deformations was
introduced in Ref.~\cite{Bizzeti04}, and the transitional nuclei $^{224,226}$%
Ra, $^{224}$Th and X(5) nuclei $^{150}$Nd, $^{152}$Sm were studied with
satisfactory results in comparison with experiment~\cite{Bizzeti08,Bizzeti10}%
. Based on this model, the stable octupole deformed nucleus $^{224}$Ra was
well described in Ref.~\cite{Bizzeti13}. More researches on the octupole
deformations and correlations can be found in Ref.~\cite{Bonatsos15} and
references therein.

Besides of the quadruple and octupole deformations, the hexadecapole
deformations are also necessary for the understanding of equilibrium shapes
and the fission process of super- and hyperdeformed nuclei~\cite%
{Aberg90,Dudek92}. The observations of $\Delta I=4$ bifurcation (also called
$\Delta I=2$ staggering) staggering phenomenon in superdeformation bands~%
\cite{Flibotte93,Cederwall94,Flibotte95,Semple96,Haslip97,Haslip982} have
aroused great enthusiasm for study of hexadecapole deformations. Many
efforts have been devoted to the subject with possible explanations given in
terms of the presence of a tetrahedral symmetry~\cite%
{Hamamoto94,Pavlichenkov95,Doenau96,Haslip981}, and the absence of any
tetrahedral symmetry~\cite%
{Guidry95,Heenen95,Luo95,Macchiavelli95,Magierski96,Mikhailov95,Pavlichenkov97,Reviol96,Toki97,Luo99}%
. The parametrization of hexadecapole deformations has been discussed in
Refs.~\cite{Nazarewicz81,Sobiczewski81,Rohozinski97}.

In real nuclei, hexadecapole deformations always coexists with quadrupole
deformations. Therefore it is natural to take the quadrupole and
hexadecapole degrees of freedom simultaneously into account~\cite%
{VanIsacker15}, especially in relation to the possible appearance of
intrinsic shapes with tetrahedral or octahedral symmetry. The tetrahedral
and octahedral shapes have been predicted by the realistic mean field
calculations~\cite{Dudek02,Dudek06}, their experimental identification in
medium and heavy mass nuclei is an open problem of current interest.
Recently, the tetrahedral symmetry has been found in the light nucleus $%
^{16} $O~\cite{Bijker14}.

From the preceding analysis, we know that the quadrupole, octupole, and
hexadecapole deformations have occurred in real nuclei, and produced
significant effects to nuclear properties. Hence, it is interesting to
discuss the collective motion for any deformed system. In the paper, we
present a general formalism describing the collective motion for any
deformed system. Firstly, we give the classical Hamiltonian of collective
motion in laboratory system, then transform it into a body-fixed frame to
separate vibrations, rotation, and the coupling between them. Under the
condition of decoupling approximation, we derive out the quantized
Hamiltonian operator. As examples, we calculate the rotational spectra for
some special octupole and hexadecapole deformed systems, and analyze the
properties of rotational spectra and their dependence on deformation.

\section{The classical theory of collective motion for any deformed system}

To describe the collective motion for any deformed system, we expand the
surface radius of the system as
\begin{equation}
R\left( \vartheta ,\varphi \right) =R_{0}\left[ 1+\sum_{lm}\alpha
_{lm}Y_{lm}\left( \vartheta ,\varphi \right) \right] ,  \label{radius}
\end{equation}%
where $\alpha _{lm}$ present the deformations deviating from the spherical
shape in Laboratory frame with the relation $\alpha _{lm}^{\ast }=\left(
-\right) ^{m}\alpha _{l,-m}$, and $R_{0}$ is the equilibrium radius. When $%
\alpha _{lm}$ are regarded as variables, the Hamiltonian describing
collective motion is obtained in the following:
\begin{equation}
H=T+V,  \label{hamilton}
\end{equation}%
where the kinetic energy is expressed as
\begin{equation}
T=\frac{1}{2}\sum_{lm}B_{l}\left\vert \dot{\alpha}_{lm}\right\vert ^{2},
\label{kinetic}
\end{equation}%
and the potential energy takes the form
\begin{equation}
V=\frac{1}{2}\sum_{lm}C_{l}\left\vert \alpha _{lm}\right\vert ^{2}.
\label{potential}
\end{equation}%
Here, $B_{l}$ and $C_{l}$ are respectively the parameters reflecting the
vibrational strength and the elastic coefficient against deformation. In the
Hamiltonian $H$, vibrations and rotation are entangled together. It is
difficult to study collective motion by using this $H$. In order to separate
vibrations and rotation from $H$, it is necessary to transform the variables
in the collective Hamiltonian from Laboratory frame ($K$-system) to
body-fixed frame ($K^{\prime }$-system) by rotation, which is defined by
\begin{equation}
R(\theta _{i})={e^{-i\theta _{2}J_{3}}e^{-i\theta _{1}J_{2}}e^{-i\theta
_{3}J_{3}},}  \label{rotation}
\end{equation}%
where $J_{1}$, $J_{2}$, and $J_{3}$ are the angular momenta along the fixed
coordinate axes ($K$-system), and $\theta _{i}=(\theta _{1},\theta
_{2},\theta _{3})$ are the Euler angles characterizing the orientation of $%
K^{\prime }$ with respect to a fixed frame of reference $K$. Through the
rotation, the variables $\alpha _{lm}$ in $K-$system can be transformed into
$K^{\prime }-$system as
\begin{equation}
\alpha _{lm}=\sum_{m^{\prime }}D_{mm^{\prime }}^{l}\beta _{lm^{\prime }},
\label{transform}
\end{equation}%
where $\beta _{lm}$ are the deformation variables in the body-fixed frame,
and $D_{mm^{\prime }}^{l}(\theta _{i})$ are the Wigner function of $\theta
_{i}$. In $D_{mm^{\prime }}^{l}(\theta _{i})$, $l$ is the angular-momentum
quantum number, $m$ and $m^{\prime }$ are the projections of angular
momentum on the laboratory fixed $z$ axis and the body-fixed $z^{\prime }$
axis, respectively.
\begin{equation}
D_{mm^{\prime }}^{l}(\theta _{i})=\left\langle lm\right\vert {e^{-i\theta
_{2}J_{3}}e^{-i\theta _{1}J_{2}}e^{-i\theta _{3}J_{3}}}{}\left\vert
lm^{\prime }\right\rangle .  \label{Dfunction}
\end{equation}%
In order to present the collective Hamiltonian using the variables ($\beta
_{lm},\theta _{i}$), we calculate the time derivative of $\alpha _{lm}$ as
\begin{equation}
\dot{\alpha}_{lm}=\sum_{m^{\prime }}\left[ D_{mm^{\prime }}^{l}(\theta _{i})%
\dot{\beta}_{lm^{\prime }}+\dot{D}_{mm^{\prime }}^{l}\left( \theta
_{i}\right) \beta _{lm^{\prime }}\right] ,  \label{alphaderivation}
\end{equation}%
where the time derivative of $D_{mm^{\prime }}^{l}(\theta _{i})$ is
presented as
\begin{equation}
\dot{D}_{mm^{\prime }}^{l}\left( \theta _{i}\right)
=-i\sum_{k}D_{mk}^{l}(\theta _{i})\left\langle lk\right\vert \vec{\omega}%
\cdot \vec{J}\left\vert lm^{\prime }\right\rangle .  \label{Dderivation}
\end{equation}%
In Eq.(\ref{Dderivation}),
\begin{eqnarray}
\omega _{1} &=&\dot{\theta}_{1}\sin \theta _{3}{-}\dot{\theta}_{2}\sin
\theta _{1}\cos \theta _{3},  \notag \\
\omega _{2} &=&\dot{\theta}_{1}\cos \theta _{3}{+}\dot{\theta}_{2}\sin
\theta _{1}\sin \theta _{3},  \notag \\
\omega _{3} &=&\dot{\theta}_{3}+\dot{\theta}_{2}\cos \theta _{1},
\label{omega}
\end{eqnarray}%
are angular velocities around the axes coincide with the body ($K^{\prime }$%
-system). Putting $\dot{\alpha}_{lm}$ into Eq.(\ref{kinetic}), the kinetic
energy splits into three parts. The first part is quadratic in $\dot{\beta}%
_{lm}$ and represents vibrations by which the body changes its shape, but
retains its orientation. The second part, quadratic in $\dot{\theta}_{i}$,
represents the rotation of the body without change of shape. The third part
contains the mixed time derivatives $\dot{\beta}_{lm}\cdot \dot{\theta}_{i}$%
, as can be shown from simple properties of the $D_{mm^{\prime }}^{l}$%
-functions and their derivatives. we thus write
\begin{equation}
T=T_{\text{vib}}+T_{\text{rot}}+T_{\text{cou}}.  \label{decomposition}
\end{equation}%
In Eq.(\ref{decomposition}), the vibrational energy
\begin{equation}
T_{\text{vib}}=\frac{1}{2}\sum_{lm}B_{l}\left\vert \dot{\beta}%
_{lm}\right\vert ^{2},  \label{Tvibration}
\end{equation}%
the rotational energy
\begin{equation}
T_{\text{rot}}=\frac{1}{2}\sum_{i,j}\mathscr{J}_{ij}\omega _{i}\omega _{j},
\label{Trotation}
\end{equation}%
with the moments of inertia
\begin{equation}
\mathscr{J}_{ij}=\frac{1}{2}\sum_{lmm^{\prime }}B_{l}\left\langle {%
lm^{\prime }}\right\vert \left\{ J_{i},\,J_{j}\right\} \left\vert
lm\right\rangle \beta _{lm}\beta _{lm^{\prime }}^{\ast },  \label{inertia}
\end{equation}%
and the coupling between vibrations and rotation
\begin{equation}
T_{\text{cou}}=\sum_{i}\omega _{i}\kappa _{i},  \label{Tcoupling}
\end{equation}%
with
\begin{equation}
\kappa _{i}=-\mathrm{Im}\sum_{lmm^{\prime }}B_{l}\left\langle {lm^{\prime }}%
\right\vert J_{i}\left\vert lm\right\rangle \dot{\beta}_{lm}\beta
_{lm^{\prime }}^{\ast }.  \label{kappa}
\end{equation}%
Here, the internal variables $\beta _{lm}$ are of complex number. For
simplicity, we introduce a set of real parameters $a_{lm}$ and $b_{lm}$ to
describe the deformations as follows:
\begin{eqnarray}
\sum_{l,m}\beta _{lm}Y_{lm}\left( \theta ,\phi \right)
&=&\sum_{l}a_{l0}Y_{l0}\left( \theta ,\phi \right)   \notag \\
&&+\sum_{l,m>0}\left[ a_{lm}Y_{lm}^{\left( +\right) }\left( \theta ,\phi
\right) +b_{lm}Y_{lm}^{\left( -\right) }\left( \theta ,\phi \right) \right] .
\label{betasurface}
\end{eqnarray}%
Here, the spherical harmonics
\begin{eqnarray}
Y_{lm}^{\left( +\right) }\left( \theta ,\phi \right)  &=&\frac{1}{\sqrt{2}}%
\left[ Y_{lm}\left( \theta ,\phi \right) +Y_{lm}^{\ast }\left( \theta ,\phi
\right) \right] ,  \notag \\
Y_{lm}^{\left( -\right) }\left( \theta ,\phi \right)  &=&\frac{1}{i\sqrt{2}}%
\left[ Y_{lm}\left( \theta ,\phi \right) -Y_{lm}^{\ast }\left( \theta ,\phi
\right) \right] .  \label{Ylmpm}
\end{eqnarray}%
From Eq.(\ref{betasurface}), we obtain
\begin{equation}
\beta _{l0}=a_{l0},\beta _{l,m}=\frac{a_{lm}-ib_{lm}}{\sqrt{2}},\beta
_{l,-m}=\left( -1\right) ^{m}\frac{a_{lm}+ib_{lm}}{\sqrt{2}},
\end{equation}%
where $m=1,2,3,\cdots ,l$. Then, the kinetic energy of vibrations in the
body-fixed frame becomes
\begin{equation}
T_{\text{vib}}=\frac{1}{2}\sum_{l}B_{l}\left[ \dot{a}_{l0}^{2}+\sum_{m>0}%
\left( \dot{a}_{lm}^{2}+\dot{b}_{lm}^{2}\right) \right] .  \label{Tvibab}
\end{equation}%
By using the relations
\begin{eqnarray}
J_{\pm }|lm\rangle  &=&\sqrt{(l{\mp }m)(l{\pm }m+1)}|l\,m\pm 1\rangle ,
\notag \\
J_{3}|lm\rangle  &=&m|lm\rangle .  \label{Jangular}
\end{eqnarray}%
Here $J_{\pm }=J_{1}{\pm }iJ_{2}$. $\kappa _{i}$ and $\mathscr{J}_{ij}$ are
expressed as the functions of the real variables $a_{lm}$ and $b_{lm}$ as
follows
\begin{eqnarray}
\kappa _{1} &=&\frac{1}{2}\sum_{l}B_{l}\bigg\{-\sqrt{2l\left( l+1\right) }%
\dot{a}_{l0}b_{l1}  \notag \\
&&+\sum_{m>0}o_{m}^{l}\left( -\dot{a}_{lm}b_{lm+1}+\dot{b}%
_{lm}a_{lm+1}\right)   \notag \\
&&+\sum_{m>0}o_{-m}^{l}\left( -\dot{a}_{lm}b_{lm-1}+\dot{b}%
_{lm}a_{lm-1}\right) \bigg\},  \label{kappa1}
\end{eqnarray}%
\begin{eqnarray}
\kappa _{2} &=&\frac{1}{2}\sum_{l}B_{l}\bigg\{\sqrt{2l\left( l+1\right) }%
\dot{a}_{l0}a_{l1}  \notag \\
&&+\sum_{m>0}o_{m}^{l}\left( \dot{a}_{lm}a_{lm+1}+\dot{b}_{lm}b_{lm+1}%
\right)   \notag \\
&&-\sum_{m>0}o_{-m}^{l}\left( \dot{a}_{lm}a_{lm-1}+\dot{b}%
_{lm}b_{lm-1}\right) \bigg\},  \label{kappa2}
\end{eqnarray}%
\begin{equation}
\kappa _{3}=\sum_{l,m>0}B_{l}m\left( a_{lm}\dot{b}_{lm}-\dot{a}%
_{lm}b_{lm}\right) ,  \label{kappa3}
\end{equation}%
\begin{eqnarray}
\mathscr{J}_{11} &=&\frac{1}{4}\sum_{l}B_{l}\bigg\{2l(l+1)a_{l0}^{2}+\sqrt{%
2l(l^{2}-1)(l+2)}a_{l0}a_{l2}  \notag \\
&&+2\sum_{m>0}\left[ l\left( l+1\right) -m^{2}\right] \left(
a_{lm}^{2}+b_{lm}^{2}\right)   \notag \\
&&+\sum_{m>0}o_{m}^{l}o_{m+1}^{l}\left( a_{lm}a_{lm+2}+b_{lm}b_{lm+2}\right)
\notag \\
&&+\sum_{m>0}o_{-m}^{l}o_{-m+1}^{l}\left(
a_{lm}a_{lm-2}+b_{lm}b_{lm-2}\right) \bigg\},  \label{J11}
\end{eqnarray}%
\begin{eqnarray}
\mathscr{J}_{22} &=&\frac{1}{4}\sum_{l}B_{l}\bigg\{2l(l+1)a_{l0}^{2}-\sqrt{%
2l(l^{2}-1)(l+2)}a_{l0}a_{l2}  \notag \\
&&+2\sum_{m>0}\left[ l\left( l+1\right) -m^{2}\right] \left(
a_{lm}^{2}+b_{lm}^{2}\right)   \notag \\
&&-\sum_{m>0}o_{m}^{l}o_{m+1}^{l}\left( a_{lm}a_{lm+2}+b_{lm}b_{lm+2}\right)
\notag \\
&&-\sum_{m>0}o_{-m}^{l}o_{-m+1}^{l}\left(
a_{lm}a_{lm-2}+b_{lm}b_{lm-2}\right) \bigg\},  \label{J22}
\end{eqnarray}%
\begin{equation}
J_{33}=\sum_{l,m>0}B_{l}m^{2}\left( a_{lm}^{2}+b_{lm}^{2}\right) ,
\label{J33}
\end{equation}%
\begin{eqnarray}
\mathscr{J}_{12} &=&\frac{1}{4}\sum_{l}B_{l}\bigg\{\sqrt{2l(l^{2}-1)(l+2)}%
a_{l0}b_{l2}  \notag \\
&&+\sum_{m>0}o_{m}^{l}o_{m+1}^{l}\left( a_{lm}b_{lm+2}-b_{lm}a_{lm+2}\right)
\notag \\
&&-\sum_{m>0}o_{-m}^{l}o_{-m+1}^{l}\left(
a_{lm}b_{lm-2}-b_{lm}a_{lm-2}\right) \bigg\},  \label{J12}
\end{eqnarray}%
\begin{eqnarray}
\mathscr{J}_{13} &=&\frac{1}{4}\sum_{l}B_{l}\bigg\{\sqrt{2l(l+1)}a_{l0}a_{l1}
\notag \\
&&+\sum_{m>0}(2m+1)o_{m}^{l}\left( a_{lm}a_{lm+1}+b_{lm}b_{lm+1}\right)
\notag \\
&&+\sum_{m>0}(2m-1)o_{-m}^{l}\left( a_{lm}a_{lm-1}+b_{lm}b_{lm-1}\right) %
\bigg\},  \label{J13}
\end{eqnarray}%
\begin{eqnarray}
\mathscr{J}_{23} &=&\frac{1}{4}\sum_{l}B_{l}\bigg\{\sqrt{2l(l+1)}a_{l0}b_{l1}
\notag \\
&&+\sum_{m>0}(2m+1)o_{m}^{l}\left( a_{lm}b_{lm+1}-b_{lm}a_{lm+1}\right)
\notag \\
&&-\sum_{m>0}(2m-1)o_{-m}^{l}\left( a_{lm}b_{lm-1}-b_{lm}a_{lm-1}\right) %
\bigg\},  \label{J23}
\end{eqnarray}%
where
\begin{equation}
o_{m}^{l}=\sqrt{\left( l-m\right) \left( l+m+1\right) },  \label{olm}
\end{equation}%
and the moments of inertia are real symmetrical: $\mathscr{J}_{ij}=%
\mathscr{J}_{ji}$. These formulas have presented a general formalism
describing the collective motion for any deformed system, where the
collective motion is treated as vibrations in the body-fixed frame ($a_{lm}$
and $b_{lm}$ vibrations), rotation of whole system about the axes of
laboratory system, and the coupling between vibrations and rotation.

The general formalism can be applied to describe the collective motion of a
classical system with any deformation. However, it should be noticed that
the variables $a_{lm}$ and $b_{lm}$ are not independent each other. Three of
them have been replaced by the Euler angles. How to remove off three
superfluous variables is a trouble problem. For the octupole and higher
multipolar deformed systems, the problem could be solved in many ways, too
many to have an obvious and natural definition of the body-fixed frame.

Some progresses have been achieved for octupole deformed system. The surface
radius expressed by the parameters $a_{3m}$ and $b_{3m}$ was
re-parameterized by a set of biharmonic coordinates~\cite%
{Rohozinski88,Rohozinski90}. In the parameterizations, the system obeys
relatively simple transformation rules under the $O_{h}$ group. Similar
parametrization was finished in Ref.~\cite{Hamamoto91}, where the intrinsic
frame was defined with four independent variables, which is a simple
combination of $a_{3m}$ and $b_{3m}$. In order to remove off the
off-diagonal elements of inertia tensor, in Ref.~\cite{Wexler99}, the
intrinsic frame was defined by the variables ($X,Y,Z,\gamma $). In
comparison with the present formalism, there exist the following relations: $%
\ $%
\begin{eqnarray}
\ \ \beta _{33} &=&\frac{1}{\sqrt{2}}a_{33}-{i}\frac{1}{\sqrt{2}}b_{33}
\notag \\
&=&\left( \cos \gamma -\frac{\sqrt{3}}{2}\sin \gamma \right) X+i\left( \cos
\gamma +\frac{\sqrt{3}}{2}\sin \gamma \right) Y,  \notag \\
\beta _{32} &=&\frac{1}{\sqrt{2}}a_{32}-{i}\frac{1}{\sqrt{2}}b_{32}=\frac{1}{%
\sqrt{2}}\sin \gamma \cdot Z\ ,  \notag \\
\beta _{31} &=&\frac{1}{\sqrt{2}}a_{31}-{i}\frac{1}{\sqrt{2}}b_{31}=\frac{%
\sqrt{5}}{2}\sin \gamma \cdot X+i\frac{\sqrt{5}}{2}\sin \gamma \cdot Y,
\notag \\
\beta _{30} &=&a_{30}=\sqrt{5}\cos \gamma \cdot Z.
\end{eqnarray}%
Namely,
\begin{eqnarray}
a_{33} &=&\left( \sqrt{2}\cos \gamma -\sqrt{3/2}\sin \gamma \right) X,
\notag \\
b_{33} &=&-\left( \sqrt{2}\cos \gamma +\sqrt{3/2}\sin \gamma \right) Y,
\notag \\
a_{32} &=&\sin \gamma \cdot Z,  \notag \\
b_{32} &=&0,  \notag \\
a_{31} &=&\sqrt{5/2}\sin \gamma \cdot X,  \notag \\
b_{31} &=&-\sqrt{5/2}\sin \gamma \cdot Y,  \notag \\
a_{30} &=&\sqrt{5}\cos \gamma \cdot Z.  \label{xyz}
\end{eqnarray}%
Putting Eqs.(\ref{xyz}) into Eqs.(\ref{J11}-\ref{J23}), for a pure octupole
deformed system, we can obtain $J_{12}=J_{13}=J_{23}=0$, and $J_{11},J_{22}$%
, and $J_{33}$ fitting the results in Ref.~\cite{Wexler99}. For example:
\begin{eqnarray}
J_{12} &=&2\sqrt{15}a_{30}b_{32}-6a_{31}b_{31}+\sqrt{15}a_{31}b_{33}-\sqrt{15%
}a_{33}b_{31}  \notag \\
&=&-6\left( \sqrt{\frac{5}{2}}\sin \gamma \cdot X\right) \left( -\sqrt{\frac{%
5}{2}}\sin \gamma \cdot Y\right)   \notag \\
&&+\sqrt{15}\left( \sqrt{\frac{5}{2}}\sin \gamma \cdot X\right) \left( -%
\sqrt{2}\cos \gamma -\sqrt{\frac{3}{2}}\sin \gamma \right) Y  \notag \\
&&-\sqrt{15}\left( \sqrt{2}\cos \gamma -\sqrt{\frac{3}{2}}\sin \gamma
\right) X\left( -\sqrt{\frac{5}{2}}\sin \gamma \cdot Y\right)   \notag \\
&=&0.
\end{eqnarray}%
Similarly, we can also reproduce the inertia tensor in Refs.~\cite%
{Bizzeti04,Bonatsos05,Bonatsos15} by a correct replacement of deformation
parameters in the present formalism.

From these discussions, we have known that there are simple relations
between the parameters in Refs.~\cite%
{Rohozinski88,Rohozinski90,Hamamoto91,Wexler99} and $\left(
a_{lm},b_{lm}\right) $ in the present formalism. Hence, the results in these
literatures~\cite{Rohozinski88,Rohozinski90,Hamamoto91,Wexler99} can be
obtained by the present formalism. Particularly, the present formalism is
appropriate to describe the collective motion for not only the systems
defined in Refs.~\cite{Rohozinski88,Rohozinski90,Hamamoto91,Wexler99} but
also those with other deformations, which is useful to investigate the
atomic nuclei with some special deformations.

In real nuclei, octupole deformations always coexist with quadrupole
deformations. Many researches~\cite%
{Nazarewicz84,Rohozinski82,Lenis06,Wexler99,Bizzeti04,Bizzeti08,Bizzeti10,Bizzeti13,Bonatsos15,Dzyublik93,Denisov95,Bonatsos05}
have been performed for the system with the coexistence of quadrupole and
octupole deformations. The present formalism is convenient to describe the
coexistence of quadrupole and octupole deformations. When the parameters
including the coexistence of quadrupole and octupole deformations are
defined properly, the Hamiltonian in Refs.~\cite%
{Nazarewicz84,Rohozinski82,Lenis06,Wexler99,Bizzeti04,Bizzeti08,Bizzeti10,Bizzeti13,Bonatsos15,Dzyublik93,Denisov95,Bonatsos05}
can be obtained using the present formalism.

Similarly, hexadecapole deformations always coexist with quadrupole
deformations in real nuclei. Many researches have been performed for the
collective motion for hexadecapole deformations coexisting with quadrupole
deformations~\cite{VanIsacker15}. Especially for the nuclei with tetrahedral
and octahedral shapes, which have been predicted by realistic mean-field
method~\cite{Dudek02,Dudek06}, and verified in recent experiment~\cite%
{Bijker14}, the present formalism is convenient to take the quadrupole and
hexadecapole degrees of freedom simultaneously into account. When $a_{lm}$
and $b_{lm}$ are re-parameterized according to the scheme in Refs.~\cite%
{Dudek02,Dudek06}, the nuclei with tetrahedral and octahedral shapes can be
studied by the present formalism. In addition, the pure hexadecapole
deformations is also concerned. In Refs.~\cite{Nazarewicz81,Sobiczewski81,Rohozinski97}, the parametrization of pure hexadecapole deformations has been discussed, and the surface
radius of the system is represented as
\begin{eqnarray}
R\left( \theta ,\phi \right)  &=&R_{0}\bigg\{1+a_{40}Y_{40}\left( \theta
,\phi \right)   \notag \\
&&+\sum_{\mu >0}\left[ a_{4\mu }Y_{4\mu }^{\left( +\right) }\left( \theta
,\phi \right) +b_{4\mu }Y_{4\mu }^{\left( -\right) }\left( \theta ,\phi
\right) \right] \bigg\}.  \label{Rhex}
\end{eqnarray}%
Here, the definitions of $Y_{\lambda \mu }^{\left( +\right)
}\left( \theta ,\phi \right) $ and $Y_{\lambda \mu }^{\left( -\right)
}\left( \theta ,\phi \right) $ are the same as those in Eqs.(\ref{Ylmpm}). It shows that the parameters $\left( a_{40},a_{4\mu },b_{4\mu },\mu =1,2,3,4\right) $ are just some special sampling of $(a_{lm}$ and $b_{lm})$. To make the system obey
relatively simple transformation rules under the $O_{h}$ group, this set of parameters $%
\left( a_{40},a_{4\mu },b_{4\mu }\right) $ have been re-parameterized with a
set of biharmonic coordinates. As there exists a simple relationship between
these biharmonic coordinates and $\left( a_{40},a_{4\mu },b_{4\mu }\right) $%
, it is easy to give out these results in Refs.~\cite%
{Nazarewicz81,Sobiczewski81,Rohozinski97} using the present formalism. The
present formalism is appropriate for any deformed system including those
defined in Refs.~\cite%
{Nazarewicz81,Sobiczewski81,Rohozinski97,Dudek02,Dudek06}, and can be used
to explore the collective motion for the system with special shape.

The preceding formalism is suitable for a classical system. To describe the
collective motion of a quantum system like atomic nucleus, it is necessary
to quantize the collective Hamiltonian. In the following, we derive out the
quantized Hamiltonian for the collective motion with any deformation.

\section{Quantization of the classical Hamiltonian}

Considering that the internal variables $a_{lm}$ and $b_{lm}$ in the present
formalism are not independent, we need to remove off three superfluous
variables among $a_{lm}$ and $b_{lm}$ in order to quantify the collective
Hamiltonian. For a quadruple deformed system, we can regard $a_{21}$, $%
b_{21} $, and $b_{22}$ as superfluous variables. When $a_{21}$, $b_{21}$,
and $b_{22}$ are removed off, $T_{\text{cou}}$ disappears, Bohr Hamiltonian
can be obtained conveniently by a simple quantization procedure. For any
deformed system, it is difficult for us to pick out three superfluous
variables in order to remove off $T_{\text{cou}}$. Even a octupole deformed
system, a set of internal parameters that make $T_{\text{cou}}$ disappear,
is not still found up to now. Here, we adopt an approximate method to
eliminate $T_{\text{cou}}$ by freezing a part of deformation parameters.
From Eqs.(\ref{kappa1}-\ref{kappa3}), we can see, to make $T_{\text{cou}}$
disappear, there are many choices of freezing deformation parameters. In the
case of freezing the least deformation parameters, the most appropriate
choice of freezing deformation parameters is that $a_{l0},a_{l2},a_{l4},%
\cdots ,a_{l,l\,or\,l-1}$ are reserved and the rest are removed off. Then,
the kinetic energy becomes
\begin{eqnarray}
T &=&\frac{1}{2}B_{2}\left( \dot{a}_{20}^{2}+\dot{a}_{22}^{2}\right) +\frac{1%
}{2}B_{3}\left( \dot{a}_{30}^{2}+\dot{a}_{32}^{2}\right)  \notag \\
&&+\frac{1}{2}B_{4}\left( \dot{a}_{40}^{2}+\dot{a}_{42}^{2}+\dot{a}%
_{44}^{2}\right) +\cdots  \notag \\
&&+\frac{1}{2}\left( \mathscr{J}_{1}\omega _{1}^{2}+\mathscr{J}_{2}\omega
_{2}^{2}+\mathscr{J}_{3}\omega _{3}^{2}\right) ,
\end{eqnarray}%
with the moments of inertia
\begin{eqnarray}
\mathscr{J}_{1} &=&\frac{1}{4}\sum_{l}B_{l}\bigg\{2l(l+1)a_{l0}^{2}+\sqrt{%
2l(l^{2}-1)(l+2)}a_{l0}a_{l2}  \notag \\
&&+\sum_{m=2,4}^{l\,\text{or}\,l-1}\left(
o_{m}^{l}o_{m+1}^{l}a_{lm}a_{lm+2}+o_{-m}^{l}o_{-m+1}^{l}a_{lm}a_{lm-2}%
\right)  \notag \\
&&+2\sum_{m=2,4}^{l\,\text{or}\,l-1}\left[ l(l+1)-m^{2}\right] a_{lm}^{2}%
\bigg\},
\end{eqnarray}%
\begin{eqnarray}
\mathscr{J}_{2} &=&\frac{1}{4}\sum_{l}B_{l}\bigg\{2l(l+1)a_{l0}^{2}-\sqrt{%
2l(l^{2}-1)(l+2)}a_{l0}a_{l2}  \notag \\
&&-\sum_{m=2,4}^{l\,\text{or}\,l-1}\left(
o_{m}^{l}o_{m+1}^{l}a_{lm}a_{l,m+2}+o_{-m}^{l}o_{-m+1}^{l}a_{lm}a_{l,m-2}%
\right)  \notag \\
&&+2\sum_{m=2,4}^{l\,\text{or}\,l-1}\left[ l(l+1)-m^{2}\right] a_{lm}^{2}%
\bigg\},
\end{eqnarray}%
\begin{equation}
\mathscr{J}_{3}=\sum_{l}B_{l}\sum_{m=2,4}^{l\,\text{or}\,l-1}m^{2}a_{lm}^{2}.
\end{equation}%
To obtain a quantized Hamiltonian, we write the kinetic energy as
\begin{equation}
T=\frac{1}{2}g_{ij}\dot{q}_{i}\dot{q}_{j},
\end{equation}%
where $q_{i}=a_{20},a_{22},a_{30},a_{32},a_{40},a_{42},a_{44},\cdots ,\phi
_{1},\phi _{2},\phi _{3}$. The metric matrix $G$ is diagonal, i.e.,
\begin{equation}
G=\left[
\begin{array}{llllllll}
B_{2} & B_{2} & B_{3} & B_{3} & \cdots & \mathscr{J}_{1} & \mathscr{J}_{2} & %
\mathscr{J}_{3}%
\end{array}%
\right] ,
\end{equation}%
its determinant
\begin{equation}
g=\det G=B_{2}^{2}B_{3}^{2}\cdots \mathscr{J}_{1}\mathscr{J}_{2}\mathscr{J}%
_{3}.
\end{equation}%
As $G$ is diagonal, $G^{-1}$ can be calculated easily. By using a usual
quantized procedure, the quantized kinetic operator is obtained as
\begin{eqnarray}
T &=&-\frac{\hbar ^{2}}{2}\frac{1}{\sqrt{g}}\frac{\partial }{\partial q_{i}}%
G_{ij}^{-1}\sqrt{g}\frac{\partial }{\partial q_{j}}  \notag \\
&=&-\frac{\hbar ^{2}}{2B_{i}}\frac{1}{\sqrt{\mathscr{J}_{1}\mathscr{J}_{2}%
\mathscr{J}_{3}}}\frac{\partial }{\partial q_{i}}\sqrt{\mathscr{J}_{1}%
\mathscr{J}_{2}\mathscr{J}_{3}}\frac{\partial }{\partial q_{i}}  \notag \\
&&+\sum\limits_{i=1}^{3}\frac{R_{i}^{2}}{2\mathscr{J}_{i}},
\end{eqnarray}%
where $q_{i}=a_{20},a_{22},a_{30},a_{32},a_{40},a_{42},a_{44},\cdots $, and $%
R_{i}=-i\hbar \frac{\partial }{\partial \phi _{i}}\;(i=1,2,3)$ are the
components of angular momentum in the intrinsic frame. In the kinetic energy
operator, the rotational part has been separated. If the only quadruple,
octupole, and hexadecapole deformations are considered, with the
transformations
\begin{eqnarray}
a_{20} &=&\beta _{2}\cos \gamma _{2},  \notag \\
a_{22} &=&\beta _{2}\sin \gamma _{2},  \notag \\
a_{30} &=&\beta _{3}\cos \gamma _{3},  \notag \\
a_{32} &=&\beta _{3}\sin \gamma _{3},  \notag \\
a_{40} &=&\beta _{4}\cos \gamma _{4},  \notag \\
a_{42} &=&\beta _{4}\cos \delta _{4}\sin \gamma _{4},  \notag \\
a_{44} &=&\beta _{4}\sin \delta _{4}\sin \gamma _{4},
\end{eqnarray}%
the quantized kinetic operator in the curve coordinates is obtained.

For a pure quadruple deformed system, we obtain immediately
\begin{eqnarray}
T_{2} &=&-\frac{\hbar ^{2}}{2B_{2}}\left( \frac{1}{\beta _{2}^{4}}\frac{%
\partial }{\partial \beta _{2}}\beta _{2}^{4}\frac{\partial }{\partial \beta
_{2}}+\frac{1}{\beta _{2}^{2}\sin 3\gamma _{2}}\frac{\partial }{\partial
\gamma _{2}}\sin 3\gamma _{2}\frac{\partial }{\partial \gamma _{2}}\right)
\notag \\
&&+\sum_{i=1}^{3}\frac{R_{i}^{2}}{2\mathscr{J}_{i}},  \label{T2}
\end{eqnarray}%
with
\begin{equation}
\mathscr{J}_{i}=4B_{2}\beta _{2}^{2}\sin ^{2}\left( \gamma _{2}-i\frac{2\pi
}{3}\right) ,\text{ \ \ \ }\left( i=1,2,3\right) .
\end{equation}%
$T_{2}$ is the kinetic energy operator in Bohr Hamiltonian.

For a pure octuple deformed system, we obtain
\begin{eqnarray}
T_{3} &=&-\frac{\hbar ^{2}}{2B_{3}}\left( \frac{1}{\beta _{3}^{4}}\frac{%
\partial }{\partial \beta _{3}}\beta _{3}^{4}\frac{\partial }{\partial \beta
_{3}}+\frac{1}{\beta _{3}^{2}w\left( \gamma _{3}\right) }\frac{\partial }{%
\partial \gamma _{3}}w\left( \gamma _{3}\right) \frac{\partial }{\partial
\gamma _{3}}\right)  \notag \\
&&+\sum_{i=1}^{3}\frac{R_{i}^{2}}{2\mathscr{J}_{i}},  \label{T3}
\end{eqnarray}%
with
\begin{equation}
w\left( \gamma _{3}\right) =\sin \gamma _{3}\sqrt{9-21\sin ^{2}\gamma
_{3}+16\sin ^{4}\gamma _{3}},
\end{equation}%
and the moments of inertia
\begin{eqnarray}
\mathscr{J}_{1} &=&B_{3}\beta _{3}^{2}\left[ 1+8\sin ^{2}\left( \gamma
_{3}+\gamma _{0}\right) \right] ,  \notag \\
\mathscr{J}_{2} &=&B_{3}\beta _{3}^{2}\left[ 1+8\sin ^{2}\left( \gamma
_{3}-\gamma _{0}\right) \right] ,  \notag \\
\mathscr{J}_{3} &=&4B_{3}\beta _{3}^{2}\sin ^{2}\gamma _{3},
\end{eqnarray}
where $\gamma _{0}=$arc$\tan \sqrt{5/3}$.

For a pure hexadecapole deformed system, we obtain
\begin{eqnarray}
T_{4} &=&-\frac{\hbar ^{2}}{2B_{4}}\bigg[\frac{1}{\beta _{4}^{5}}\frac{%
\partial }{\partial \beta _{4}}\beta _{4}^{5}\frac{\partial }{\partial \beta
_{4}}  \notag \\
&&+\frac{1}{\beta _{4}^{2}\sin \gamma _{4}w\left( \gamma _{4},\delta
_{4}\right) }\frac{\partial }{\partial \gamma _{4}}\sin \gamma _{4}w\left(
\gamma _{4},\delta _{4}\right) \frac{\partial }{\partial \gamma _{4}}  \notag
\\
&&+\frac{1}{\beta _{4}^{2}\sin ^{2}\gamma _{4}w\left( \gamma _{4},\delta
_{4}\right) }\frac{\partial }{\partial \delta _{4}}w\left( \gamma
_{4},\delta _{4}\right) \frac{\partial }{\partial \delta _{4}}\bigg]  \notag
\\
&&+\sum\limits_{i=1}^{3}\frac{R_{i}^{2}}{2\mathscr{J}_{i}},  \label{T4}
\end{eqnarray}%
with
\begin{equation}
w\left( \gamma _{4},\delta _{4}\right) =\sqrt{\mathscr{J}_{1}^{\prime }%
\mathscr{J}_{2}^{\prime }\mathscr{J}_{3}^{\prime }},
\end{equation}%
and the moments of inertia
\begin{equation}
\mathscr{J}_{i}=B_{4}\beta _{4}^{2}\mathscr{J}_{i}^{\prime }\text{ , \ \ \ \
}\left( i=1,2,3\right) .
\end{equation}%
Here
\begin{eqnarray}
\mathscr{J}_{1}^{\prime } &=&10+3\sqrt{5}\cos \delta _{4}\sin 2\gamma _{4}
\notag \\
&&+\left( 3\cos 2\delta _{4}+\sqrt{7}\sin 2\delta _{4}-5\right) \sin
^{2}\gamma _{4},  \notag \\
\mathscr{J}_{2}^{\prime } &=&10-3\sqrt{5}\cos \delta _{4}\sin 2\gamma _{4}
\notag \\
&&+\left( 3\cos 2\delta _{4}+\sqrt{7}\sin 2\delta _{4}-5\right) \sin
^{2}\gamma _{4},  \notag \\
\mathscr{J}_{3}^{\prime } &=&\left( \cos ^{2}\delta _{4}+4\sin ^{2}\delta
_{4}\right) \sin ^{2}\gamma _{4}.
\end{eqnarray}

From Eq.(\ref{T2}), we notice that the 4th power of $\beta _{2}$ appears in
the first term of the kinetic energy. The same case also appears in Eq.(\ref%
{T3}) for $\beta _{3}$. Different from Eqs.(\ref{T2}) and (\ref{T3}), the
5th power of $\beta _{4}$ appears in the first term of the kinetic energy.
It is because the power of $\beta _{i}\left( i=2,3,4\right) $ appearing in
the first term of the kinetic energy depends on the number of degrees of
freedom. For $T_{2}$ and $T_{3}$, only two deformation variables ($%
a_{20},a_{22}$) and ($a_{30},a_{32}$) are taken into account, while for $%
T_{4}$, three deformation variables $\left( a_{40},a_{42},a_{44}\right) $
are taken into account.

\section{The rotational spectra for some special deformed systems}

In the preceding section, we have derived the quantized kinetic operator for
multipolar deformed systems, including the quadruple, octupole, and
hexadecapole deformed systems. When the potential against deformation is
included, the quantized Hamiltonian operator describing multipolar deformed
system is obtained. The Hamiltonian can be used to study the collective
motion of a quantum system with multipolar deformations. As the Hamiltonian
is complicated, here we do not discuss in details solution of the general
Hamiltonian. Follow Davydov's assumption, we regard the deformation
variables as parameters, and investigate the rotation of multipolar deformed
systems, which is very interesting to study the rotational spectra in atomic
nuclei.

In order to obtain the rotational spectra for some special deformed systems,
we introduce the axially symmetrical spheroidal wave functions
\begin{equation}
\left\vert IK\pm \right\rangle =\sqrt{\frac{2I+1}{16\pi ^{2}\left( 1+\delta
_{K0}\right) }}\left[ D_{MK}^{I}\pm \left( -1\right) ^{I}D_{M,-K}^{I}\right],
\end{equation}%
as bases in calculating the energy spectra of rotational Hamiltonian. As $%
P\left\vert IK\pm \right\rangle =\pm \left\vert IK\pm \right\rangle $, where
$P$ is parity operator, we choose $\left\vert IK,+\right\rangle $ as bases
for the positive parity states, and $\left\vert IK,-\right\rangle $ as bases
for the negative parity states.

By using Eqs.(\ref{rotation}), (\ref{Dfunction}), and (\ref{Jangular}), we
obtain the following equations:
\begin{eqnarray}
R_{1}^{2}\left\vert IK\pm \right\rangle &=&\frac{1}{2}\left[ I\left(
I+1\right) -K^{2}\right] \left\vert IK\pm \right\rangle  \notag \\
&&+\frac{1}{4}o_{K}^{I}o_{K+1}^{I}\left\vert I,K+2\pm \right\rangle  \notag
\\
&&+\frac{1}{4}o_{K-1}^{I}o_{K-2}^{I}\left\vert I,K-2\pm \right\rangle , \\
R_{2}^{2}\left\vert IK\pm \right\rangle &=&\frac{1}{2}\left[ I\left(
I+1\right) -K^{2}\right] \left\vert IK\pm \right\rangle  \notag \\
&&-\frac{1}{4}o_{K}^{I}o_{K+1}^{I}\left\vert I,K+2\pm \right\rangle  \notag
\\
&&-\frac{1}{4}o_{K-1}^{I}o_{K-2}^{I}\left\vert I,K-2\pm \right\rangle , \\
R_{3}^{2}\left\vert IK\pm \right\rangle &=&K^{2}\left\vert IK\pm
\right\rangle ,
\end{eqnarray}%
where $R_{1}$, $R_{2}$, and $R_{3}$ are the rotational operators around the
first, second, and third axis in the body-fixed frame, respectively. The
expression of $o_{K}^{I}$ is the same as $o_{m}^{l}$ in Eq.(\ref{olm}). With
the relations, the matrix elements of the rotational operator are obtained
as
\begin{eqnarray}
&&\left\langle IK^{\prime }\right\vert \sum\limits_{i=1}^{3}\frac{R_{i}^{2}}{%
2\mathscr{J}_{i}}\left\vert IK\right\rangle  \notag \\
&=&\frac{1}{4}\left( \frac{1}{\mathscr{J}_{1}}+\frac{1}{\mathscr{J}_{2}}%
\right) I\left( I+1\right) \delta _{K^{\prime }K}  \notag \\
&&+\frac{1}{2}\left( \frac{1}{\mathscr{J}_{3}}-\frac{1}{2\mathscr{J}_{1}}-%
\frac{1}{2\mathscr{J}_{2}}\right) K^{2}\delta _{K^{\prime }K}  \notag \\
&&+\frac{1}{8}\left( \frac{1}{\mathscr{J}_{1}}-\frac{1}{\mathscr{J}_{2}}%
\right) \sqrt{1+\delta _{K0}}o_{K}^{I}o_{K+1}^{I}\delta _{K^{\prime }K+2}
\notag \\
&&+\frac{1}{8}\left( \frac{1}{\mathscr{J}_{1}}-\frac{1}{\mathscr{J}_{2}}%
\right) \sqrt{1+\delta _{K^{\prime }0}}o_{K-1}^{I}o_{K-2}^{I}\delta
_{K^{\prime }K-2}.  \label{matrix}
\end{eqnarray}

By using Eq.(\ref{matrix}), we can study the collective rotation for the
system with the deformations $a_{l0},a_{l2},a_{l4},\cdots ,a_{l,l\,or\,l-1}$%
. Here, we do not focus on the full spectrum of a deformed nucleus with
dominant quadrupole deformation. We are only concerned about rotational
spectra for the system with pure octupole or hexadecapole deformations.
Although octupole or hexadecapole deformations always coexist with
quadrupole deformations in real nuclei, our studies can provide some
information on rotational spectra for octupole and hexadecapole deformed
systems, which are helpful to know the properties of atomic nuclei with
octupole or hexadecapole deformations coexisting with quadrupole
deformations.

Considering that the most interesting rotational spectra are those with the
lowest $K$, we have calculated the rotational spectra with the lowest $K$
for the octupole and hexadecapole deformed systems. In Fig.~1, we have shown
the variation of rotational spectra with $\gamma _{3}$ for an octupole
deformed system. For simplicity, we take $2^{+}$ state as an example to
analyze the relationship between the level energy and $\gamma _{3}$
deformation. For $2^{+}$ state, there are two levels. The first (lowest) $%
2^{+}$ level is denoted by red solid line and the second $2^{+}$ level by
red dash line. With the change of $\gamma _{3}$, the first $2^{+}$ level
varies slowly. In the vicinity of $\gamma _{3}=0^{\circ }$, the first $2^{+}$
level appears a little decreasing with the decreasing $\gamma _{3}$, while
that appears a little increasing with the increasing $\gamma _{3}$ closing
to $\gamma _{3}=90^{\circ }$. In the range of $\gamma _{3}=20^{\circ }$ and $%
70^{\circ }$, the energy of the first $2^{+}$ level is nearly a constant.
The same phenomena also appear in the first $3^{+}$ level, the first $4^{+}$
level, the first $5^{+}$ level, and the first $6^{+}$ level. For all these
levels with the same angular momentum and parity, the lowest level is
insensitive to $\gamma _{3}$. Different from these lowest levels, the second
and third levels in every angular momentum and parity go to infinity with $%
\gamma _{3}$ going to zero. With the increasing of $\gamma _{3}$, the second
and third levels appear valleys, i.e., metastable states, which may be the
isomers of $\gamma _{3}$ deformation. When $\gamma _{3}=\gamma _{0}$, the
second and/or third levels appear peaks, i.e., $\gamma _{3}$ unstable
states. When $\gamma _{3}=90^{\circ }$, $a_{30}$ disappears, only $a_{32}$
deformation exists in the nuclei, the shape of this system possesses $T_{d}$
symmetry, the rotational Hamiltonian is then reduced to a spherical top, so
the rotational levels with the same angular momentum are degenerate. In a
word, the contribution of the octupole term to the spectrum is smooth for
the lowest band, while it becomes irregular for higher bands. In real
nuclei, this contribution from octupole term will be added to the dominant
quadrupole contribution, thus it will most probably result to some small
deviations from quadrupole spectrum. But, the character of octupole spectrum
can reflect the information on the properties of real nuclei with octupole
deformations coexisting with the quadrupole deformations.

\begin{figure}[tbp]
\includegraphics[width=9.5cm]{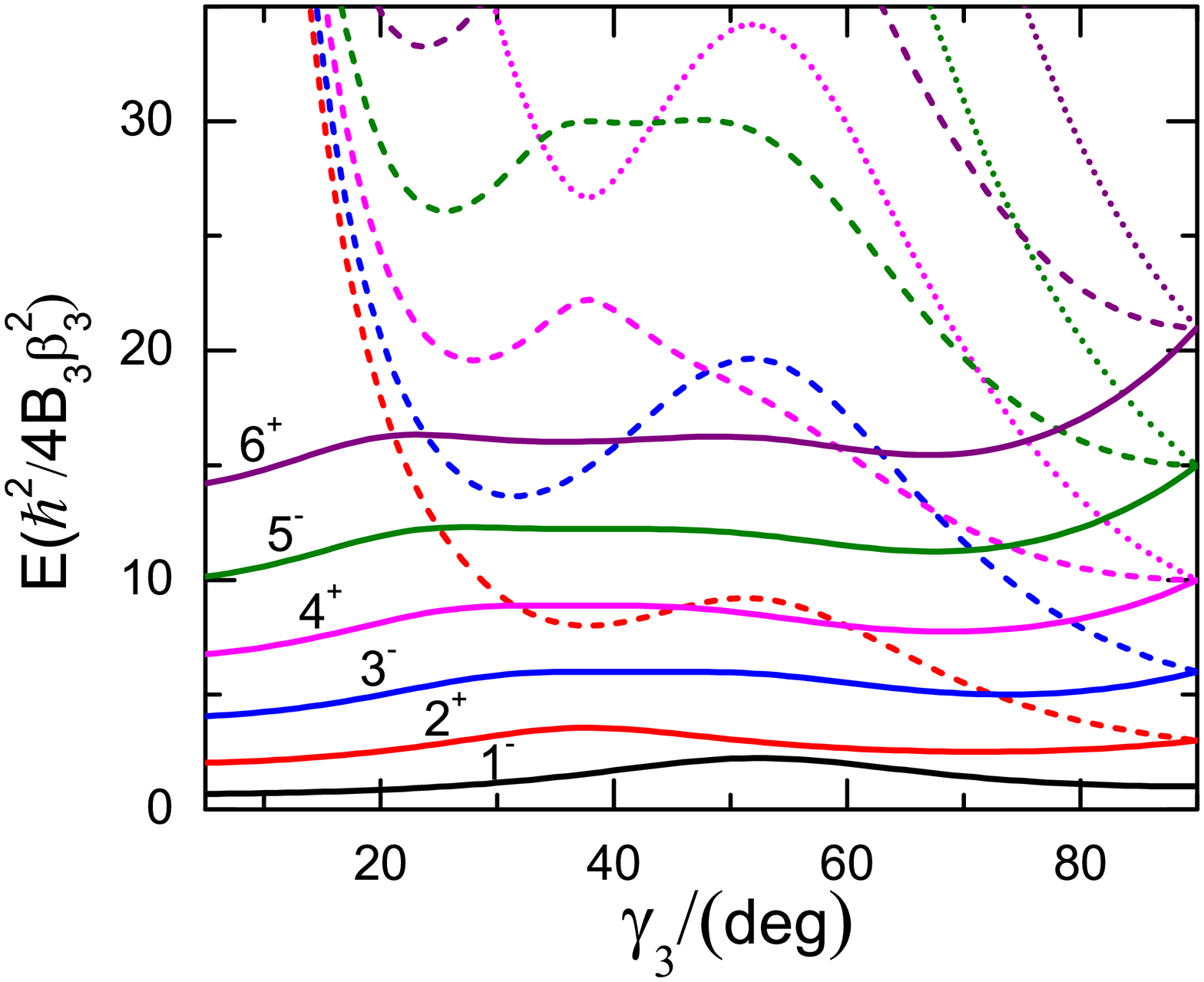}
\caption{(Color online) Evolution of rotational spectra to $\protect\gamma %
_{3}$ for an octupole deformed system. Here, the solid, dash, and dot lines
with the same color represent respectively the first, second, and third
energy levels for these states with the same angular momentum and parity.}
\end{figure}

Besides of the octupole deformed system, we have also calculated the
rotational spectra for a hexadecapole deformed system. In Fig.~2, we
demonstrate the evolution of rotational spectra to $\gamma _{4} $ with $%
\delta _{4}$ fixed to $0$, i.e., only $a_{40}$ and $a_{42}$ deformations
under consideration. Over the range of $\gamma _{4}$, the lowest levels of
even angular momentum states are almost independent of $\gamma _{4}$. Only
in the vicinity of $\gamma _{4}=0^{\circ }$ and $\gamma _{4}=90^{\circ }$,
these levels appear a little decreasing or increasing with $\gamma _{4}$.
However for these levels corresponding to the odd and higher even angular
momentum states, their energies are sensitive to $\gamma _{4}$. Similar to
that of octupole deformation, these levels go to infinity when $\gamma _{4}$
goes to $0^{\circ }$. With the increasing of $\gamma _{4}$, these levels
drop quickly, but not monotonously, appear valley: metastable state, which
may be the isomer of $\gamma _{4}$ deformation, and peak: unstable state.
When $\gamma _{4}$ is added to $90^{\circ }$, $a_{40}$ disappears, all the
rotational levels become relatively low, which implies that it is relatively
easy to appear $a_{42}$ deformation in real nuclei.
\begin{figure}[tbp]
\includegraphics[width=9.5cm]{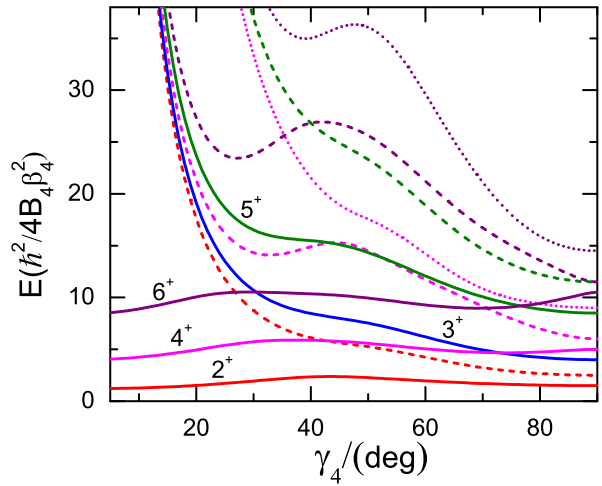}
\caption{(Color online) Evolution of rotational spectra to $\protect\gamma %
_{4}$ for a hexadecapole deformed system ($\protect\delta $ is fixed to $%
0^{\circ }$). Here, the solid, dash, and dot lines with the same color
represent respectively the first, second, and third energy levels for these
states with the same angular momentum and parity.}
\end{figure}

When $\delta _{4}$ is fixed to $45^{\circ }$, the rotational spectra varying
with $\gamma _{4}$ is displayed in Fig.~3. Except for the lowest levels of $%
2^{+}$ and $4^{+}$ states, the other levels depend remarkably on $\gamma
_{4} $. Only in the vicinity of $\gamma _{4}=0^{\circ }$, the lowest levels
of even angular momentum states keep a good structure of rotational spectra,
while the other levels go to infinity. With the increasing of $\gamma _{4}$,
these levels corresponding to the odd and higher even angular momentum
states change dramatically. In the region around $\gamma _{4}=30^{\circ }$
and $\gamma _{4}=90^{\circ }$, all the levels are relatively low. In the
other region, except for the lowest levels of $2^{+}$ and $4^{+}$, the other
levels are too high so that it is difficult to appear these levels in real
nuclei. Furthermore, a sharp peak appears in these levels, which corresponds
to $\gamma _{4}$ unstable state. As the peak is too high, it is impossible
to appear the $\gamma _{4}$ unstable state in real nuclei, which is
different from that in Fig. 2.
\begin{figure}[tbp]
\includegraphics[width=9.5cm]{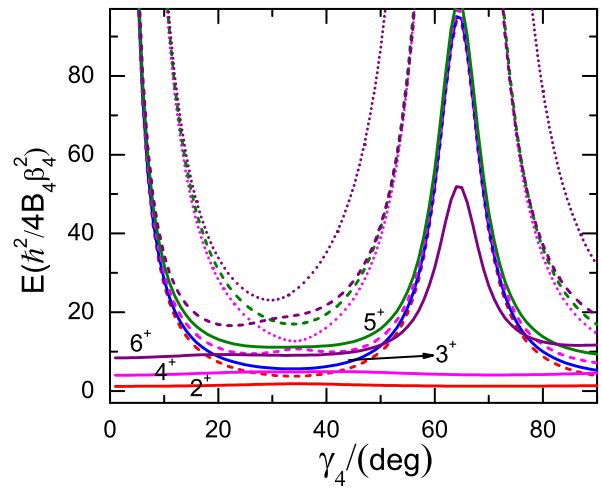}
\caption{(Color online) The same as Fig.2, but $\protect\delta_4 $ is fixed
to $45^{\circ }$.}
\end{figure}

In Fig.~4, we show the variation of rotational spectra with $\gamma _{4}$
for $\delta _{4}=90^{\circ }$. In the case, only $a_{40}$ and $a_{44}$
deformations are concerned, the shape of system possesses $D_{4h}$ symmetry
and the corresponding moments of inertia $\mathscr{J}_{1}=\mathscr{J}_{2}$.
From Fig.~4, we can see that there exists a critical point of $\gamma _{4}$
deformation ($\gamma _{4}^{c}\approx 40.2^{\circ }$). In the point, $%
\mathscr{J}_{1}=\mathscr{J}_{2}=\mathscr{J}_{3}$, the rotational Hamiltonian
is reduced to a spherical top, the rotational levels with the same angular
momentum are degenerate. When $\gamma _{4}<\gamma _{4}^{c}$, the lowest
levels of even angular momentum states form a good rotational spectrum
although the energies of these levels increase with the increasing $\gamma
_{4}$. However for the odd angular momentum states, their energies go to
infinity when $\gamma _{4}$ goes to $0$. The same case also appears in the
second and third levels of even angular momentum states. This means that it
is difficult to appear the rotational states with odd angular momentum or
the excited states with even angular momentum in the vicinity of $\gamma
_{4}=0^{\circ }$. When $\gamma _{4}>\gamma _{4}^{c}$, the energies of all
the levels increase with the increasing $\gamma _{4}$, which shows that it
is more unstable for the nuclei with a larger $\gamma _{4}$ deformation.
\begin{figure}[tbp]
\includegraphics[width=9.5cm]{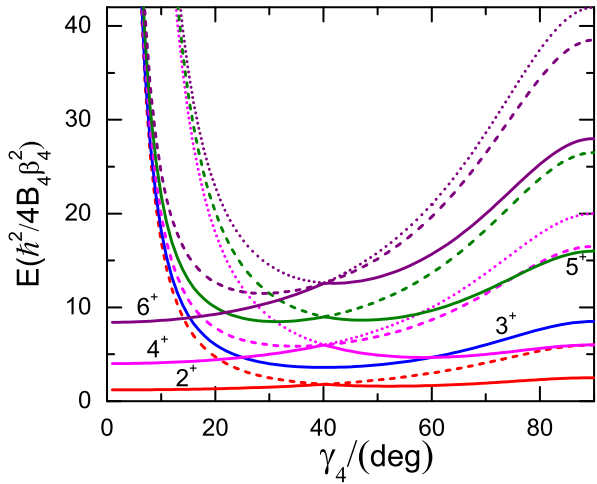}
\caption{(Color online) The same as Fig.2, but $\protect\delta_4 $ is fixed
to $90^{\circ }$.}
\end{figure}

Over Figs.~2-4, we can see that the contributions of hexadecapole
deformations to the lowest band are regular, while those to higher bands are
irregular. In real nuclei, these contributions from hexadecapole
deformations will be added to those from the dominant quadrupole
deformations, and will bring a bit of deviations from the energy spectrum of
qudrupole deformations. But, the feature reflecting hexadecapole
deformations will be reserved, which is useful to know the properties of
real nuclei with hexadecapole deformations coexisting with the quadrupole
deformations.

\section{Conclusions}

Based on Bohr model, we have presented a general formalism describing the
collective motion for any deformed system, in which the collective
Hamiltonian is expressed as vibrations in the body-fixed frame, rotation of
whole system around the laboratory frame, and coupling between vibrations
and rotation. Under the condition of decoupling approximation, we have
derived the quantized Hamiltonian operator. Based on the operator, we have
calculated the rotational energy for some special octupole and hexadecapole
deformed systems, and shown their dependencies on deformation. In the
octupole deformed nuclei, for these states with the same angular momentum
and parity, the lowest level is insensitive to $\gamma _{3}$, and all the
lowest levels form a regular rotational spectrum. Different from the lowest
levels, the higher levels depend remarkably on $\gamma _{3}$. In the
vicinity of $\gamma _{3}=0^{\circ }$, these higher levels go to infinity.
With the increasing of $\gamma _{3}$, these levels drop quickly, but not
monotonously. There appear peak (unstable state) and valley (metastable
state) in these levels over the range of $\gamma _{3}$. These metastable
states may form the isomers of $\gamma _{3}$ deformation. The similar case
also appears in the hexadecapole deformed system with $\delta _{4}=0^{\circ
} $. The lowest levels of even angular momentum states are almost
independent of $\gamma _{4}$ and form a regular rotational band. For the odd
and higher even angular momentum states, the corresponding levels are
sensitive to $\gamma _{4}$. They go to infinity closing to $\gamma
_{4}=0^{\circ }$, and decline fast with the increasing $\gamma _{4}$.
Similarly, there appear $\gamma _{4}$ unstable and metastable states in the
range of $\gamma _{4}$. For the hexadecapole deformations with $\delta _{4}$
fixed to $45^{\circ }$ and $90^{\circ }$, the lowest levels of even angular
momentum states form regular rotational spectra in the vicinity of $\gamma
_{4}=0^{\circ }$. With the increasing of $\gamma _{4}$, these levels for the
odd and higher even angular momentum states change dramatically. These show
that the octupole and/or hexadecapole contributions to the lowest band are
regular, while those to higher bands are dramatic. In real nuclei, these
contributions will be added to a dominant quadrupole contribution, and
produce some small influences on the energy spectrum of quadrupole
deformations. Nevertheless, these features reflecting octupole and
hexadecapole deformations are helpful to understand the properties of real
nuclei with octupole and/or hexadecapole coexisting with the quadrupole
deformations.

\section{Acknowledgments}

This work was partly supported by the National Natural Science Foundation of
China under Grants No. 11175001, and No. 11205004; the Program for New
Century Excellent Talents in University of China under Grant No.
NCET-05-0558; the Excellent Talents Cultivation Foundation of Anhui Province
under Grant No. 2007Z018; the Natural Science Foundation of Anhui Province
under Grant No. 11040606M07; and the 211 Project of Anhui University.

\end{document}